\documentclass[letterpaper]{article}

\usepackage{aaai}
\usepackage{times}
\usepackage{helvet}
\usepackage{courier}
\usepackage{graphicx}
\usepackage{multirow}
\usepackage{amsmath}
\usepackage{pdfpages}
\usepackage{floatflt}
\usepackage{url}
\PassOptionsToPackage{hyphens}{url}

\DeclareMathOperator*{\argmin}{arg\,min}

\begin{document}
\title{Scalable Privacy-Compliant Virality Prediction on Twitter}
\author{Damian Konrad Kowalczyk\\
Microsoft Development Center Copenhagen\\
Kanalvej 7, 2800 Kgs. Lyngby, Denmark\\
\texttt{dakowalc@microsoft.com}
\And
Jan Larsen\\
DTU Compute, Matematiktorvet 303B\\
2800 Kgs. Lyngby, Denmark\\
\texttt{janla@dtu.dk}
}
\maketitle
\begin{abstract}
\begin{quote}
The digital town hall of Twitter becomes a preferred medium of communication for individuals and organizations across the globe. Some of them reach audiences of millions, while others struggle to get noticed. Given the impact of social media, the question remains more relevant than ever: how to model the dynamics of attention in Twitter. Researchers around the world turn to machine learning to predict the most influential tweets and authors, navigating the volume, velocity, and variety of social big data, with many compromises. In this paper, we revisit content popularity prediction on Twitter. We argue that strict alignment of data acquisition, storage and analysis algorithms is necessary to avoid the common trade-offs between scalability, accuracy and privacy compliance. We propose a new framework for the rapid acquisition of large-scale datasets, high accuracy supervisory signal and multilanguage sentiment prediction while respecting every privacy request applicable. We then apply a novel gradient boosting framework to achieve state-of-the-art results in virality ranking, already before including tweet's visual or propagation features. Our Gradient Boosted Regression Tree is the first to offer explainable, strong ranking performance on benchmark datasets. Since the analysis focused on features available early, the model is immediately applicable to incoming tweets in 18 languages.
\end{quote}
\end{abstract}

\section{Introduction and motivation}

\begin{quote}
The role of the social and professional networks in the spread and acceptance of innovations, knowledge, business practices, products, behavior, rumors, and memes, is a much-studied problem in social sciences, marketing and economics. Online environments like Twitter, offer an unprecedented opportunity to track such phenomena. Consequently, a staggering number of studies focus on social spreading, asking for example why can some messages reach millions of individuals, while others struggle to get noticed. \cite{barabasi2016network}
\end{quote}

\noindent The knowledge discovery process, however, is becoming even more tangled with the arrival of social big data. 700 million tweets have been posted on the day of writing this introduction. The volume, velocity, and variety of mostly unstructured information even from a single social network are evolving at an extremely fast pace. From an engineering and data science perspective, near real-time analysis via online services and algorithms scalable in-memory are required, and demand substantial computational resources. Scientific endeavors to date offer progress toward specific subtasks of social network analysis (SNA) yet data collection and privacy compliance remain among the biggest challenges in extracting knowledge \cite{Bello-Orgaz2016b}. Arguably the most significant among them is privacy \cite{Sapountzi2018}. The social nature of nodes in these networks makes data subjective to many privacy concerns and laws. The new European General Data Protection Regulation (GDPR and ISO/IEC 27001) in force since May 25th, 2018 makes SNA and black-box approaches (like deep neural networks) more difficult to use in business, requiring the results to be retraceable (explainable) on demand \cite{Holzinger2017}. In machine learning, explainable (compliant) real-time analysis is often at odds with predictive accuracy. In social popularity prediction, some of the best results today are achieved using deep neural networks, difficult to interpret \cite{Wang2018a} or data modalities time-consuming to acquire \cite{Firdaus2016}. Modeling popularity relies on a precise count of responses (subject to privacy requests, i.e., retweets in virality prediction) which exposes them further. Accuracy in such studies depends on processing documents no longer available, while privacy compliance requires removing them. Ensuring accurate and explainable analysis via quality of the data and methods, while respecting user privacy, remain conflicting goals and open research issues individually. \\\\ In this work we argue that significant advancement in SNA requires avoiding such trade-offs and addressing all the above issues simultaneously. We draw inspiration from multiple disciplines, to challenge state of the art in content virality prediction on Twitter. We propose a framework which to the best of our knowledge, is the first one that satisfies the properties of model preserving and privacy-compliant simultaneously. We use it to train a scalable and explainable model, and are the first to achieve strong \cite{cohen1988spa} virality ranking performance on multiple benchmark datasets.

\section{Related work}
\subsection{Social big data analysis before GDPR}
Social big data has become essential for various distributed services, applications, and systems \cite{Peng2018}, enabling event detection \cite{Dong2015}, sentiment analysis \cite{Feldman:2013:TAS:2436256.2436274}, popularity prediction \cite{Wu2015}, natural language processing, finding influential bloggers, personalized recommendation \cite{Gan2018}, online advertising, viral marketing, opinion leader detection etc. Computational and storage requirements of such applications have led to cloud scale reinvention of data storage and processing technologies. New tools are constantly emerging to replace the conventional non-effective ones, and a hybrid of techniques \cite{Kaisler2013,Gandomi2015} is now a requirement to extract value from the social big data. \cite{Sheela2016} proposes a solution based on Hadoop technology and a Naive Bayes classification for sentiment analysis of tweets. The sentiment analysis in performed in MapReduce layer and results stored in distributed NO-SQL data-base. \cite{Huang2014} uses Lucene indexing with full-text searching ability on top of Hadoop for spectral clustering, to detect Twitter communities during the Hurricane Sandy disaster. In our work we pursue close alignment of data acquisition and analysis algorithms, with the strict constraints of storage and time, to accommodate both user-generated content (UGC) and privacy requests, arriving at high volume and velocity. Instead of perturbing or anonymizing the data, sensitive or deleted information is permanently eliminated from storage and subsequent analysis.
\subsection{Content popularity prediction}
Social network influence can be defined as the ability of a user to spread information in the network \cite{Pezzoni2013}, with the retweet count assumed as a measure of a tweet’s popularity. One common challenge for content-based popularity prediction is the 140-character constraint imposed by Twitter, making it difficult to identify and extract predictive features \cite{Can2013}. \cite{tan2014effect} showed that carefully crafted wording of the message could help propagate the tweets better, but there's much more to UGC than the caption. \cite{Ishiguro2012,Wang2018a} demonstrate social-oriented features were the best performers to predict image popularity on Twitter. \cite{McParlane2014} utilized textual, visual, and social cues to predict the image popularity on Flickr. \cite{Wang2018a} proposed a joint-embedding neural network combining the same cues to rival state-of-the-art methods. Recurrent and Deep Neural Networks advance feature extraction from high-dimensional unstructured data (i.e., image attachments), however due to low explainability also introduce a major drawback for critical decision-making processes (with recent advances by \cite{Samek2017}).  In this study, we prioritize explainable methods in application to structured data. 
\cite{Pezzoni2013,Kwak2010a,Cha2010} demonstrate relationships between the number of followers of Twitter users and their influence on information spreading. Ranking users by the number of followers is found to perform similarly to PageRank \cite{Kwak2010a}. \cite{Pezzoni2013} models the probability to be retweeted by a power law function. \cite{Palovics2013a} have used an explainable Random Forrest classifier to predict a range of the logarithm of the retweets volume. He demonstrates the predictive value of user features (e.g., count of followers), network features, and the popularity of hashtags included. \cite{Bunyamin2016} provide a comparison of learning methods and features, regarding retweet prediction accuracy and feature importance. They find Random Forests to achieve the best performance in binary classification of retweetability and highlight the value of author features: number of times the user is listed by other users, number of followers and the average number of tweets posted per day. \cite{Nesi2018} uses recursive partitioning trees to achieve 0.682 classification accuracy on a large topical dataset, albeit using features unavailable early (favorites count) or anymore (local publication time) challenging both scalability and reproducibility. \cite{Hansen2011c} investigated the features of tweets contributing to retweetability and is the first to explore the impact of negative sentiment in diffusion of news on Twitter. We follow \cite{Hansen2011c} to consider affect in our model. Substantial gains are seen when including network features extracted from the content graph formed by retweets, or relationship graph formed by "friendships". The document level subgraphs to inform prediction are often acquired via real-time monitoring of the diffusion process. \cite{Zaman2010} predicted the popularity of a tweet through the time-series path of its retweets, using a Bayesian probabilistic model. \cite{Wang2018a} uses preconditioned recurrent neural network to model the temporal diffusion, and shows SOTA ranking performance of 0.366 on benchmark datasets. \cite{Ahmed2013} used temporal evolution patterns to predict the popularity of online UGC. \cite{Cheng2014} use temporal and structural features to predict the cascades of photo shares on Facebook. \cite{DBLP:journals/corr/ZhaoEHRL15} model the retweeting cascades as a self-exciting point process. \cite{Firdaus2016} argues that determining the topic of interest of a user based on his past tweets might boost predictive accuracy. \cite{Peng2011} studied retweet network propagation trends using conditional random fields, demonstrating gains in accuracy when considering social relationships and retweet history. Access to subgraphs on the author or even document level is however strictly limited by social networks, thus leveraging tweets’ (early) performance, authors’ relationships, preferences or retweet history is prohibitive for a scalable, near real-time prediction on a single tweet. 

\noindent In this study we seek to maximize virality ranking performance. We follow \cite{Wang2018a} to approach the problem as Poisson regression, and \cite{Hansen2011c} to consider tweet sentiment in prediction. However, in the contrast to prior work, we don't sacrifice scalability or privacy compliance, nor rely on available retweet count for ground truth.

\begin{figure*}[t] 
\centering
\includegraphics[width=\textwidth]{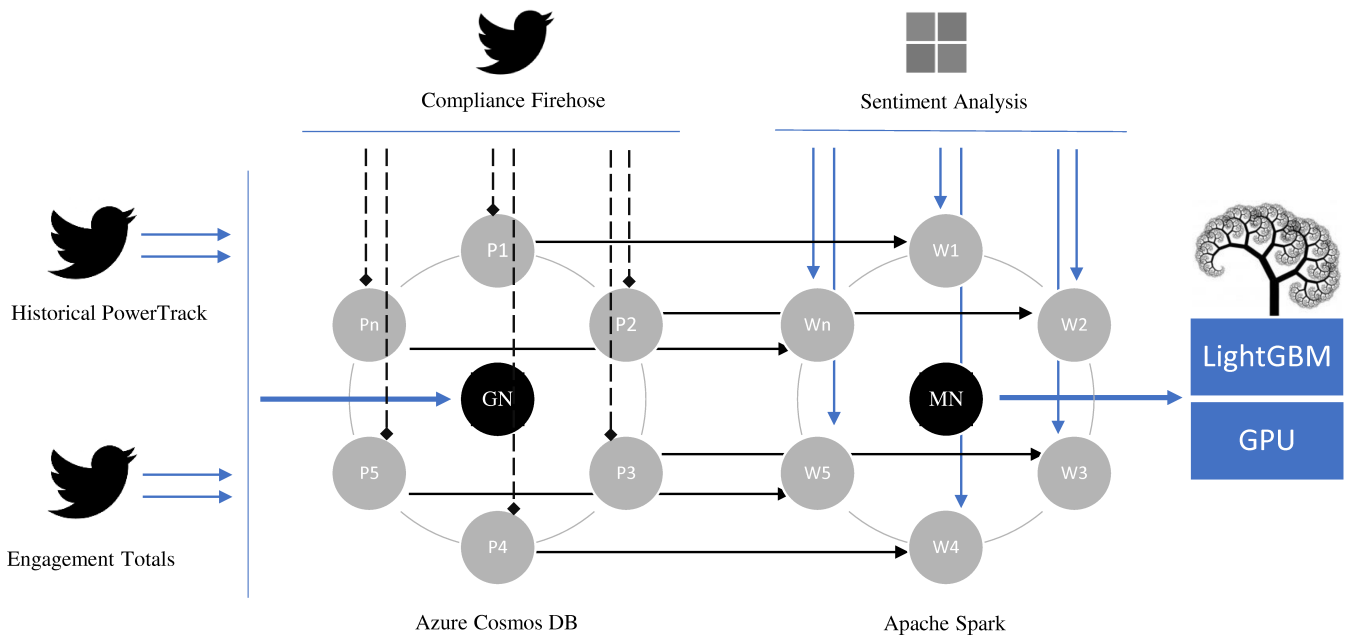}
\caption{Solution overview, including data acquisition, storage and analysis components. Cosmos DB gateway node GN orchestrates indexing of Twitter’s historical data to partitions P, for simultaneous feature extraction by Spark worker nodes W, before aggregation by master node MN for GPU accelerated predictive analysis.}
\label{fig:Solution}
\end{figure*}

\section{Solution overview}

\subsection{Data acquisition}
We use Twitter’s Historical APIs to acquire datasets of tweets for training and validation against other studies. In contrast to sampling Twitter’s x-hose, predominant in prior work, we apply Twitter’s PowerTrack search rules, to formulate and collect entire datasets retroactively. The documents are then stored in a globally distributed NO-SQL database, hosted by Microsoft Azure. The data remains online, exposed to every privacy request applicable. 

\subsection{Privacy compliant storage}
Data analyzed in this study is publicly available during collection. Exactly how much of it remains public, changes rapidly afterwards. Account removal, suspension, or deleting of a single tweet render affected content unavailable for analysis in a privacy-compliant way. Users exercise their right to be forgotten at an unprecedented rate. We consume an average of 4,000 of such requests per second via Twitter’s Compliance Firehose API and apply to our storage simultaneously with analysis. For perspective, the average rate of new tweets published today is 8,000/s. To support this velocity and rapid feature extraction for dependent analysis we choose Azure Cosmos DB as the persistent data store.      
\subsection{High accuracy labels}
In the contrast to prior work, we do not rely on available retweet count for training supervision. Twitter's Engagement Totals API is called during data collection, to retrieve the number of retweets and favorites ever registered for the tweet (including those deleted shortly after). This enables our data collection effort to focus on unique content only, reducing the document volume required for the task (and proportional compliance responsibility) by more than half, while ensuring 100\% accuracy of the supervisory signal. 
\subsection{Sentiment analysis}
To compute document sentiment, we adopt Text Analytics API from Microsoft Cognitive Services \cite{TextAnalytics}, a collection of readily consumable ML algorithms in the cloud. At the time of this study, the service supports 18 languages: English, Spanish, Portuguese, French, German, Italian, Dutch, Norwegian, Swedish, Polish, Danish, Finnish, Russian, Greek, Turkish, Arabic, Japanese and Chinese. The service is for-profit and continuously improving (changing) over time, which might challenge reproduction. To address this, we share the score of each document.
\subsection{Compute}
We conduct an in-memory analysis of entries no longer personally identifiable. This prevents fragmentation of sensitive data outside of the central store exposed to user privacy requests. Instead of anonymizing the datasets, sensitive or deleted information is eliminated from storage and future analysis as soon as the request from the user is processed by the social media platform.
We dedicate an Apache Spark cluster to data preprocessing and analysis. Spark is efficient at iterative computations and is thus well-suited for the development of large-scale machine learning applications \cite{Meng:2016:MML:2946645.2946679}. Communication performance between Spark and our privacy-compliant Cosmos DB enables feature extraction at rates exceeding 65,000 tweets per second. The resulting in-memory dataset is then aggregated by the Spark master node, equipped with Tesla K80 GPUs (Graphics Processing Units) for predictive analysis and model tuning. We choose LightGBM framework to train our Gradient Boosted Regression Tree and explain the choice in the following section.
\section{Data collection}
\begin{table*}[t!]
\centering
\resizebox{\textwidth}{!}{%
\begin{tabular}{c|ccccccc}
\hline
\textbf{Dataset}              & \textbf{Timeframe} & \textbf{Months} & \textbf{Language} & \textbf{With images only} & \textbf{Total} & \textbf{Unique tweets (acquired)} & \textbf{Never retweeted} \\ \hline
\textbf{MBI (Cappallo, 2016)} & 2013.02-2013.03    & 2               & English           & TRUE                      & 2,724,764      & 1,319,288                         & 1,042,411                \\
\textbf{T2015 (Wang, 2018)}   & 2015.11-2016.04    & 6               & English           & TRUE                      & 9,025,826      & 2,804,153                         & 2,106,475                \\
\textbf{T2016 (Wang, 2018)}   & 2016.10-2015.12    & 3               & English           & TRUE                      & 8,469,016      & 2,736,600                         & 2,088,377                \\
\textbf{T16-BIO}              & 2015.06-2017.06    & 12              & Multi (18x)       & FALSE                     & 27,032,417     & 14,788,552                        & 12,809,021               \\
\textbf{T2017-BIO}            & 2017.01-2018-02    & 14              & Multi (18x)       & FALSE                     & 19,850,448     & 9,719,264                         & 8,774,009                \\ \hline
\end{tabular}%
}
\caption{Datasets acquired}
\label{datasets}
\end{table*}

We use the new framework to build multiple datasets across different time periods for training and evaluation of our models (Table 1)

\subsubsection{Benchmark datasets}
We acquire three benchmark datasets MBI, T2015 and T2016 (with a total of 6,860,041 unique tweets) to enable comparison with the work of \cite{Mazloom2016,McParlane2014,Khosla2014,Cappallo2015,Wang2018a}. The datasets match the same filters, as applied before (e.g., timeframe, language or presence of image attachment) yet result in higher volume. We follow \cite{Wang2018a,Cappallo2015} to split the tweets into 70\% training, 10\% validation, and 20\% test sets respectively.
\subsubsection{Twitter 2017}
For the general multilanguage model, we have collected 10 million unique tweets and used 9.7M of them for predictive analysis, after applying privacy requests. The dataset has been downsampled from the entire Twitter 2017 volume to 18 languages supported by the sentiment scoring service, then using Twitter PowerTrack’s sample and bio operators, to manage the volume without sacrificing our model’s generalization capability over the full year. 
\subsection{Sentiment score and all-time totals}
Retweet counts, favorite counts, and sentiment scores were collected for ca. 30 million unique tweets, simultaneously with applying privacy requests. It is worth noting that 85\% of unique tweets acquired had never been retweeted.

\subsection{Feature selection}
Multiple features have been extracted from the rich Twitter metadata, to capture what is being said (content), by who (author), when (temporal) and how (sentiment). Table 2 describes selected features and their Pearson correlation coefficient with the logarithm of retweet count in T2017-BIO. Only the information available at the time of acquisition or immediately after is considered, to maximize the scalability of the solution. Specifically, we do not consider the early performance of the tweet (i.e., retweet or favorite counts received) or image-based features at this point. Some authors (e.g., celebrities) receive more attention than others despite low activity. We calculate the two author ratio features in an attempt to isolate such examples. Number of attachments (like hashtags, mentions, URLs, images, symbols and videos) compete for viewer’s atten-tion with the original 140-character body of the tweet, and their total count is also considered. Finally, we log-transform selected author features (e.g. author's favorite and listed counts) due to power-law distribution \cite{Can2013}.

\begin{table}[t]
\resizebox{\columnwidth}{!}{%
\begin{tabular}{lllr}
\hline
\multicolumn{1}{l}{\textbf{Modality}}  & \textbf{Feature}    & \textbf{Type} & \textbf{Pearson} \\ \hline
\multirow{7}{*}{\textbf{(A) Author}}   & followersCount      & ordinal       & 0.205920          \\
                                       & friendsCount        & ordinal       & 0.082779         \\
                                       & accountAgeDays      & ordinal       & 0.020379         \\
                                       & statusesCount       & ordinal       & -0.001455        \\
                                       & actorFavoritesCount & ordinal       & 0.029914         \\
                                       & actorListedCount    & ordinal       & 0.221067         \\
                                       & actorVerified       & categorical   & 0.202722         \\ \hline
\multirow{6}{*}{\textbf{(C) Content}}  & attachmentCount     & ordinal       & 0.085333         \\
                                       & mentionCount        & ordinal       & -0.006590         \\
                                       & hashtagsCount       & ordinal       & 0.104335         \\
                                       & mediaCount          & ordinal       & 0.147623         \\
                                       & urlCount            & ordinal       & 0.082549         \\
                                       & isQuote             & categorical   & 0.061915         \\ \hline
\multirow{2}{*}{\textbf{(L) Language}} & languageIndex       & categorical   & 0.005199         \\
                                       & sentimentValue      & continuous    & 0.059863         \\ \hline
\multirow{5}{*}{\textbf{(T) Temporal}} & postedHour          & ordinal       & 0.016639         \\
                                       & postedDay           & ordinal       & -0.000963        \\
                                       & postedMonth         & ordinal       & -0.004129        \\
                                       & postedDayTime       & categorical   & 0.016639         \\
                                       & postedWeekDay       & categorical   & -0.001002        \\ \hline
\end{tabular}%
}
\caption{Feature summary}
\label{my-label}
\end{table}

\section{Methodology}
We consider the problem of predicting the scale of retweet cascade for a given tweet based on data modalities available immediately after its delivery. The author features are used together with the content, language, and temporal to predict the number of future retweets. In this study, we assume the future retweet count \(r\) of a tweet follows Poisson distribution: 
\begin{equation}
P(R=r\mid\lambda)=\frac{e^{-\lambda}\lambda^{-r}}{r!}
\end{equation}
where the latent variable \(\lambda \in R^+ \) defines the mean and variance of the distribution, and maximize the Poisson log-likelihood given a collection of N training tuples of tweets \(t_i\) and their retweet counts \(r_{gt,i}\)

\begin{equation}
    \theta^*=\argmin_\theta\frac{1}{N}\sum[{r_{gt,i}\ln{\lambda(t_i)}+\lambda(t_i)}]
\end{equation}
where \(\theta\) contains all parameters of the proposed model.

\subsection{Gradient Boosted Regression Tree}
GBRT is a tree ensemble algorithm which builds one regression tree at a time by fitting the residual of the trees that preceded it. With our twice-differentiable loss function, denoted as:
\begin{equation}
    L_{\rm Poisson}(r_{gt},t)=r_{gt}\ln\lambda(t)+\lambda(t)
\end{equation}
GBRT minimizes the loss function (regularization term omitted for simplicity):
\begin{equation}
    L=\sum_{i=1}^NL_{\rm Poisson}(r_{gt,i},F(t_i))
\end{equation}
with a function estimation F(t) represented in an additive form:
\begin{equation}
    F(t)=\sum^T_{m=1}f_m(t)
\end{equation}
where each \(F_m(t)\) is a regression tree and \(T\) is the number of trees. GBRT learns these regression trees in an incremental way: at \(m\)-stage, fixing the previous \(m-1\) trees when learning the \(m\)-th trees. To construct the \(m\)-th tree, GBRT minimizes the following loss:
\begin{equation}
    L_m=\sum_{t=1}^NL_{\rm Poisson}(r_{gt,i},F_{m-1}(t_i)+f_m(t_i))
\end{equation}
where \(F_{m-1}\left(t\right)=\sum_{k}^{m-1}{f_k\left(t\right)}\). The optimization problem (6) can be solved by Taylor expansion of the loss function:

\begin{equation}
\begin{split}
L_m\approx\bar{L}_m=\sum^N_{i=0}[L_{\rm Poisson}(r_{gt,i},F_{m-1}(t_i)) \\ +\nabla_if_m(t_i)+\frac{\nabla^2_i}{2}f^2_m(t_i)]
\end{split}
\end{equation}
with the gradient and Hessian defined as: 
\begin{equation}
\begin{split}
    \nabla_i=\frac{\partial{L_{\rm Poisson}(r_{gt,i},F(t_i))}}{\partial{F(t_i)}}\mid{F(t_i)=F_{m-1}(t_i)}\\
    \nabla^2_i=\frac{\partial{L^2_{\rm Poisson}(r_{gt,i},F(t_i))}}{\partial^2{F(t_i)}}\mid{F(t_i)=F_{m-1}(t_i)}
\end{split}
\end{equation}
We train our GBRT by minimizing \(\bar{L}_m\) which is equivalent to minimizing: 
\begin{equation}
    \min_{f\in{F}}\sum^N_{i=1}{\frac{\nabla^2_i}{2}(f_m(t_i)+\frac{\nabla_i}{\nabla^2_i})^2}
\end{equation}
This approach is vulnerable to overdispersion and power-law distribution, characterizing the retweet count. In extreme cases where Hessian is nearly zero (9) approaches positive infinity. To safeguard the optimization, we cap each tree’s weight estimation at 1.5 and follow \cite{Can2013} to use total retweet count as ground-truth after log-transformation:

\begin{equation}
    r_{gt}=\ln(r_{total}+1)
\end{equation}

\subsection{Gradient Boosting Framework}
LightGBM \cite{Ke2017} implementation of GBDT is chosen for the task, due to distinctive techniques applicable. Experiments on multiple public datasets show that Gradient-based One-Side Sampling (GOSS) and Exclusive Feature Bundling (EFB) can accelerate the training process by over 20 times while achieving almost the same accuracy \cite{Ke2017}. Most of all, LightGBM implements a novel histogram-based algorithm to approximately find the best splits which is highly scalable on GPUs \cite{Zhang2017}. The framework allows us to explore substantially larger hyperparameter space during cross-validation. Finally, LightGBM offers good accuracy with integer-encoded categorical features by applying \cite{WalterD.Fisher1958} to find the optimal split over categories. This often performs better than one-hot encoding and enables treating more features as categorical while avoiding dimensionality explosion.

\section{Experiments}

We exercise gradient boosted Poisson regression in experiments organized by datasets, to tune and compare our approach against recent state-of-the-art methods, before attempting to generalize the prediction across topics and cultures in the multilingual extended timeframe study.

\subsection{Evaluation metrics}
We compute the Spearman Rho ranking coefficient, to measure our model’s ability to rank the content by expected popularity. Interpretation of this coefficient is domain specific, with guidelines for social/behavioral sciences proposed by \cite{cohen1988spa}. SpearmanR from SciPy version 1.4.0 is used to ensure tie handling. We did not find this concern expressed in prior work. 	The p-value for all reported Spearman results is \(p<0.001\)

Relative and absolute measures of fit: \(R^2\), and RMSE are chosen for optimization, to penalize large error higher (i.e. when underestimating highly viral content or vice-versa). The mean-absolute-percentage-error (MAPE) is computed due to popularity in previous studies \cite{Wang2018a}, but not considered for tuning. We dispute MAPE’s value relative to above when fitting asymmetric, zero-inflated distribution of the dependent variable (like retweet count). It is undefined for the majority of examples (Table 1), which never receive a retweet and penalizes errors for least retweeted higher.

\subsection{Validation on benchmark datasets}
We begin with evaluation of our multimodal GBRT against previous state-of-the-art methods. For a fair comparison, we use Poisson regression on the joint author, content and temporal features (ACT), before including sentiment (ACTL). Table \ref{benchmark} demonstrates that our proposed model achieves substantially higher ranking performance, compared to other content-based methods, already before considering image and propagation modalities. Using more advanced feature representations, sentiment score and high accuracy ground-truth, we outperform the state-of-the-art by more than 37\% on multiple datasets.

\begin{table}[ht]
\resizebox{\columnwidth}{!}{%
\begin{tabular}{llll|lll}
\hline
\textbf{Method}      & \multicolumn{3}{c|}{\textbf{SpearmanR}}                              & \multicolumn{3}{c}{\textbf{MAPE}}                                    \\ \cline{2-7} 
                     & \textbf{MBI}                       & \textbf{T2015} & \textbf{T2016} & \textbf{MBI}                       & \textbf{T2015} & \textbf{T2016} \\ \cline{2-7} 
McParlene*           & 0.188                              & 0.269          & 0.257          & 0.093                              & 0.121          & 0.137          \\
Khosla*              & 0.185                              & 0.273          & 0.254          & 0.097                              & 0.103          & 0.124          \\
Cappallo*            & 0.189                              & 0.265          & 0.258          & 0.089                              & 0.095          & 0.119          \\
Mazloom*             & 0.190                               & 0.287          & 0.262          & 0.073                              & 0.097          & 0.117          \\
Wang*                & 0.229                              & 0.358          & 0.350           & 0.057                              & 0.084          & 0.103          \\
\textbf{Ours (ACT)}  & \textbf{0.322}                     & \textbf{0.498} & \textbf{0.503} & \text{0.247}                     & \text{0.266} & \text{0.256} \\
\textbf{Ours (all)} & \textbf{0.323}                     & \textbf{0.499} & \textbf{0.504} & \text{0.247}                     & \text{0.266} & \text{0.255} \\ \hline
\textbf{}            & \multicolumn{3}{c|}{\textbf{\(R^2\)}}                              & \multicolumn{3}{c}{\textbf{RMSE}}                                    \\ \cline{2-7} 
                     & \textbf{MBI}                       & \textbf{T2015} & \textbf{T2016} & \textbf{MBI}                       & \textbf{T2015} & \textbf{T2016} \\ \cline{2-7} 
\textbf{Ours (ACT)}  & \multicolumn{1}{c}{\textbf{0.303}} & \textbf{0.417} & \textbf{0.391} & \multicolumn{1}{c}{\textbf{0.444}} & \textbf{0.553} & \textbf{0.555} \\ \hline
\end{tabular}%
}
\caption{Method performance on benchmark datasets. *measurements first published by \cite{Wang2018a}}
\label{benchmark}
\end{table}

\subsection{Multilingual, extended time-frame experiments}
We apply our method to the new T2017-BIO dataset to generalize popularity prediction across languages and time. Tweet \(t(A,C,T,L)\) includes content descriptions C, language descriptions L and is ﬁrst issued by author A, at the time T. Table 4 summarizes contributions of these modalities individually and in combination. The baseline model is trained on a single feature, most popular in literature: the count of author’s followers, notified about the tweet. 

\begin{table}[ht]
\resizebox{\columnwidth}{!}{%
\begin{tabular}{c|cccc}
\hline
\textbf{Features} & \textbf{SpearmanR} & \textbf{\(R^2\)} & \textbf{RMSE}  & \textbf{MAPE}  \\ \hline
A                 & 0.310               & 0.317              & 0.359          & 0.133          \\
C                 & \textbf{0.211}     & 0.055              & 0.422          & 0.160           \\
T                 & 0.062              & 0.001              & 0.432          & 0.171          \\
L                 & 0.164              & 0.017              & 0.430           & 0.167          \\
AC                & 0.356              & \textbf{0.396}     & 0.337          & 0.121          \\
AT                & 0.311              & 0.316              & 0.359          & 0.132          \\
AL                & 0.324              & 0.320               & 0.358          & 0.130           \\
CT                & 0.220               & 0.059              & 0.421          & 0.159          \\
CL                & 0.269              & 0.076              & 0.417          & 0.154          \\
TL                & 0.170               & 0.019              & 0.430           & 0.166          \\
ATL               & 0.324              & 0.320               & 0.358          & 0.130           \\
ACT               & 0.357              & 0.395              & 0.338          & 0.120           \\
ACL               & 0.369              & 0.399              & 0.336          & 0.119          \\
\textbf{ACTL}     & \textbf{0.369}     & \textbf{0.402}     & \textbf{0.336} & \textbf{0.118} \\ \hline
\textbf{Baseline} & \textbf{0.180}      & \textbf{0.091}     & \textbf{0.414} & \textbf{0.160}  \\ \hline
\end{tabular}%
}
\caption{Quantitative evaluation of ‘A’: actor, ‘C’: content, ‘T’: temporal, and ‘L’: language features. SpearmanR, R squared: higher is better. RMSE, MAPE: lower is better}
\label{multiexp}
\end{table}

\section{Discussion}
When prioritizing social posts by expected popularity, model's ranking performance might precede metrics of overall fit. Interpretation of Spearman ranking coefficient and \(R^2\) metrics is domain specific. For social/behavioral sciences, reaching 0.5 indicates strong correlation \cite{cohen1988spa}. The final study aimed to explore generalizability of our method over an extended time-frame and 18 languages. The relative insignificance of the Temporal modality (Table \ref{multiexp}) suggests low correlation between the time of posting and the content popularity, thereby challenging the common intuition, that posting at the time of audience’s activity helps propagating the content. We also find that content-based features alone have higher value for expected popularity ranking than the number of followers. How many people like you appears less important than what you have to say. 

Non-linear advanced ML algorithms like deep neural net-works and gradient boosted decision trees are among the most successful methods used today. The fact is often attributed to the inherent capability of discovering non-linear relationships between groups of features. It was not necessary in our study to compute e.g., all cross-products to rival state-of-the-art, and at times we have noticed a higher cumulative contribution of combined modalities over their individual gains (Table 4). 
The size of the audience immediately exposed to the tweet, measured as the count of the author’s followers, remains the single strongest predictor of tweet popularity when considered in isolation (Figure 2). The number of times an author has been listed by others, followed others or favorited other content are also among significant features, open to interpretation. Number of friends is arguably related to the diversity of content the author is exposed to. We expect the count of tweets favorited over time (i.e. age of account) to differentiate active from passive consumers. Assuming the author’s influence is measured by her capacity to spread information in the social network \cite{Pezzoni2013}, could the diversity of content actively consumed over time maximize author’s influence? We propose this hypothesis for computational social science.

\begin{figure*}[t!] 
\centering
\includegraphics[width=\textwidth]{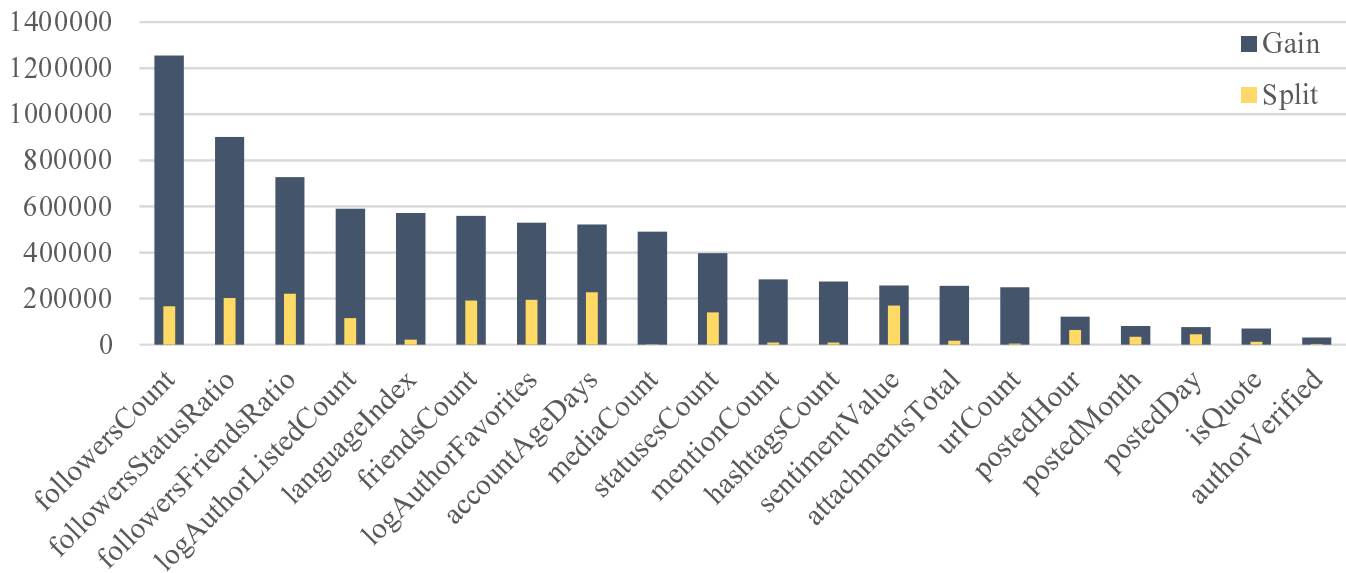}
\caption{Feature level importance}
\label{fig:importance}
\end{figure*}

\section{Conclusions and future work}
In this paper, we have studied the problem of predicting tweet popularity under scalability, explainability and privacy compliance constraints. Our method estimates the potential reach of a tweet i.e. size of retweet cascades based on modalities available immediately after document creation. We prove it is possible to rival state-of-the-art results without compromising on explainability, scalability or privacy compliance. Our Gradient Boosted Regression Tree, combining available modalities with sentiment score and high accuracy ground-truth achieves state-of-the-art results on multiple datasets and is the first to achieve strong \cite{cohen1988spa} virality ranking performance. \\\indent In the final round of experiments, we apply our method to generalize prediction across extended time-frame in 18 languages and explain the contribution of each modality. 
Training the final model on NVidia Tesla K80 took 10 minutes. Computing predictions for the 2 million unique tweets in the validation set, took another 45 seconds. This implies throughput of over 44,000 tweets / second, with a single GPU. Assuming incoming tweets are already vectorized, the ACT model deployed on Tesla K80 can cope with 5 (five) times today’s Twitter volume and velocity. \cite{Wang2018a} take up to 72 additional hours (after data collection) to acquire propagation features for the prediction. During that time, our model will  have predicted popularity for up to 11 billion tweets.
\subsection{Applications}  
Our model is ready for production with immediate application to social media monitoring. The proposed framework is extendable to other data modalities (e.g. visual) and other methods (e.g. deep neural networks)
\subsubsection{Storage} Our privacy compliant storage solution is immediately applicable to data collection and analysis from other social networks exposing privacy signal (e.g. Tumblr and WordPress, with privacy requests available as “compliance interactions” from DataSift). 
\subsubsection{Compute} Our solution to focus analysis on temporary in-memory samples, created ad-hoc for every iteration, from a single central persistent storage to receive compliance requests, is applicable to any social network sourced data.
\subsubsection{High-accuracy labels} Our solution to rely on dedicated APIs for high accuracy labels (i.e. count of retweets, replies or likes/favorites ever registered) instead of error prone counting or crawling used in prior work, is immediately applicable to Instagram, Tumblr and Facebook Pages.
\subsubsection{Multimodal GBM} Our histogram-based gradient boosted regression approach is immediately applicable to Instagram, Tumblr and Facebook Pages.

\subsection{Acknowledgements}
This project is supported by Microsoft Development Center Copenhagen and the Danish Innovation Fund, Case No. 5189-00089B. We would like to thank Charlotte Mark, Lars Kai Hansen, Joerg Derungs, Uffe Kjall, Petter Stengard, Piotr Madejski and Tomasz Janiczek. Any opinions, findings, conclusions or recommendations expressed in this material are those of the authors and do not necessarily reflect those of the sponsors.

\bibliographystyle{aaai}
\bibliography{virality}

\clearpage
\begin{thebibliography}{}

\bibitem[\protect\citeauthoryear{Ahmed, Spagna, and Huici}{2013}]{Ahmed2013}
Ahmed, M.; Spagna, S.; and Huici, F.
\newblock 2013.
\newblock {A Peek into the Future : Predicting the Evolution of Popularity in
  User Generated Content}.
\newblock In {\em Proceedings of the sixth ACM international conference on Web
  search and data mining}.

\bibitem[\protect\citeauthoryear{Barabasi and
  Posfai}{2016}]{barabasi2016network}
Barabasi, A.-L., and Posfai, M.
\newblock 2016.
\newblock {\em Network science}.
\newblock Cambridge: Cambridge University Press.

\bibitem[\protect\citeauthoryear{Bello-Orgaz, Jung, and
  Camacho}{2016}]{Bello-Orgaz2016b}
Bello-Orgaz, G.; Jung, J.~J.; and Camacho, D.
\newblock 2016.
\newblock {Social big data: Recent achievements and new challenges}.
\newblock {\em Information Fusion}.

\bibitem[\protect\citeauthoryear{Bunyamin and Tunys}{2016}]{Bunyamin2016}
Bunyamin, H., and Tunys, T.
\newblock 2016.
\newblock {A Comparison of Retweet Prediction Approaches: The Superiority of
  Random Forest Learning Method}.
\newblock {\em TELKOMNIKA (Telecommunication Computing Electronics and
  Control)} 14(3):1052.

\bibitem[\protect\citeauthoryear{Can, Oktay, and Manmatha}{2013}]{Can2013}
Can, E.~F.; Oktay, H.; and Manmatha, R.
\newblock 2013.
\newblock {Predicting retweet count using visual cues}.
\newblock In {\em Proceedings of the 22nd ACM international conference on
  Conference on information {\&} knowledge management - CIKM '13}.

\bibitem[\protect\citeauthoryear{Cappallo, Mensink, and
  Snoek}{2015}]{Cappallo2015}
Cappallo, S.; Mensink, T.; and Snoek, C.~G.
\newblock 2015.
\newblock {Latent Factors of Visual Popularity Prediction}.
\newblock In {\em Proceedings of the 5th ACM on International Conference on
  Multimedia Retrieval - ICMR '15}.

\bibitem[\protect\citeauthoryear{Cha \bgroup et al\mbox.\egroup
  }{2010}]{Cha2010}
Cha, M.; Haddadi, H.; Benevenuto, F.; and Gummadi, K.~P.
\newblock 2010.
\newblock {Measuring User Influence in Twitter: The Million Follower Fallacy}.
\newblock In {\em ICWSM 10}.

\bibitem[\protect\citeauthoryear{Cheng \bgroup et al\mbox.\egroup
  }{2014}]{Cheng2014}
Cheng, J.; Adamic, L.~A.; Dow, P.~A.; Kleinberg, J.; and Leskovec, J.
\newblock 2014.
\newblock {Can Cascades be Predicted?}

\bibitem[\protect\citeauthoryear{Cohen}{1988}]{cohen1988spa}
Cohen, J.
\newblock 1988.
\newblock {\em {Statistical Power Analysis for the Behavioral Sciences}}.
\newblock Lawrence Erlbaum Associates.

\bibitem[\protect\citeauthoryear{Dong \bgroup et al\mbox.\egroup
  }{2015}]{Dong2015}
Dong, X.; Mavroeidis, D.; Calabrese, F.; and Frossard, P.
\newblock 2015.
\newblock {Multiscale event detection in social media}.
\newblock {\em Data Mining and Knowledge Discovery}.

\bibitem[\protect\citeauthoryear{Feldman}{2013}]{Feldman:2013:TAS:2436256.2436274}
Feldman, R.
\newblock 2013.
\newblock Techniques and applications for sentiment analysis.
\newblock {\em Commun. ACM} 56(4):82--89.

\bibitem[\protect\citeauthoryear{Firdaus, Ding, and
  Sadeghian}{2016}]{Firdaus2016}
Firdaus, S.~N.; Ding, C.; and Sadeghian, A.
\newblock 2016.
\newblock {Retweet prediction considering user's difference as an author and
  retweeter}.
\newblock In {\em Proceedings of the 2016 IEEE/ACM International Conference on
  Advances in Social Networks Analysis and Mining, ASONAM 2016}.

\bibitem[\protect\citeauthoryear{Fisher}{1958}]{WalterD.Fisher1958}
Fisher, W.~D.
\newblock 1958.
\newblock {On Grouping For Maximum Homogeneity}.
\newblock {\em American Statistical Association Journal}.

\bibitem[\protect\citeauthoryear{Gan and Jiang}{2018}]{Gan2018}
Gan, M., and Jiang, R.
\newblock 2018.
\newblock {FLOWER: Fusing global and local associations towards personalized
  social recommendation}.
\newblock {\em Future Generation Computer Systems}.

\bibitem[\protect\citeauthoryear{Gandomi and Haider}{2015}]{Gandomi2015}
Gandomi, A., and Haider, M.
\newblock 2015.
\newblock {Beyond the hype: Big data concepts, methods, and analytics}.
\newblock {\em International Journal of Information Management}.

\bibitem[\protect\citeauthoryear{Hansen \bgroup et al\mbox.\egroup
  }{2011}]{Hansen2011c}
Hansen, L.~K.; Arvidsson, A.; Nielsen, F.~A.; Colleoni, E.; and Etter, M.
\newblock 2011.
\newblock {Good friends, bad news - Affect and virality in twitter}.
\newblock In {\em Communications in Computer and Information Science}.

\bibitem[\protect\citeauthoryear{Holzinger \bgroup et al\mbox.\egroup
  }{2017}]{Holzinger2017}
Holzinger, A.; Biemann, C.; Pattichis, C.~S.; and Kell, D.~B.
\newblock 2017.
\newblock {What do we need to build explainable AI systems for the medical
  domain?}

\bibitem[\protect\citeauthoryear{Huang \bgroup et al\mbox.\egroup
  }{2014}]{Huang2014}
Huang, Y.; Dong, H.; Yesha, Y.; and Zhou, S.
\newblock 2014.
\newblock {A Scalable System for Community Discovery in Twitter During
  Hurricane Sandy}.
\newblock In {\em 2014 14th IEEE/ACM International Symposium on Cluster, Cloud
  and Grid Computing},  893--899.
\newblock IEEE.

\bibitem[\protect\citeauthoryear{Ishiguro, Kimura, and
  Takeuchi}{2012}]{Ishiguro2012}
Ishiguro, K.; Kimura, A.; and Takeuchi, K.
\newblock 2012.
\newblock {Towards automatic image understanding and mining via social
  curation}.
\newblock In {\em Proceedings - IEEE International Conference on Data Mining,
  ICDM}.

\bibitem[\protect\citeauthoryear{Kaisler \bgroup et al\mbox.\egroup
  }{2013}]{Kaisler2013}
Kaisler, S.; Armour, F.; Espinosa, J.~A.; and Money, W.
\newblock 2013.
\newblock {Big data: Issues and challenges moving forward}.
\newblock In {\em Proceedings of the Annual Hawaii International Conference on
  System Sciences}.

\bibitem[\protect\citeauthoryear{Ke \bgroup et al\mbox.\egroup }{2017}]{Ke2017}
Ke, G.; Meng, Q.; Wang, T.; Chen, W.; Ma, W.; Liu, T.-Y.; Finley, T.; Wang, T.;
  Chen, W.; Ma, W.; Ye, Q.; and Liu, T.-Y.
\newblock 2017.
\newblock {LightGBM: A highly efficient gradient boosting decision tree}.
\newblock {\em Advances in Neural Information Processing Systems}.

\bibitem[\protect\citeauthoryear{Khosla, {Das Sarma}, and
  Hamid}{2014}]{Khosla2014}
Khosla, A.; {Das Sarma}, A.; and Hamid, R.
\newblock 2014.
\newblock {What makes an image popular?}
\newblock In {\em Proceedings of the 23rd international conference on World
  wide web - WWW '14}.

\bibitem[\protect\citeauthoryear{Kwak \bgroup et al\mbox.\egroup
  }{2010}]{Kwak2010a}
Kwak, H.; Lee, C.; Park, H.; and Moon, S.
\newblock 2010.
\newblock {What is Twitter, a social network or a news media?}
\newblock In {\em Proceedings of the 19th international conference on World
  wide web - WWW '10}.

\bibitem[\protect\citeauthoryear{Mazloom \bgroup et al\mbox.\egroup
  }{2016}]{Mazloom2016}
Mazloom, M.; Rietveld, R.; Rudinac, S.; Worring, M.; and van Dolen, W.
\newblock 2016.
\newblock {Multimodal Popularity Prediction of Brand-related Social Media
  Posts}.
\newblock In {\em Proceedings of the 2016 ACM on Multimedia Conference - MM
  '16}.

\bibitem[\protect\citeauthoryear{McParlane, Moshfeghi, and
  Jose}{2014}]{McParlane2014}
McParlane, P.~J.; Moshfeghi, Y.; and Jose, J.~M.
\newblock 2014.
\newblock {"Nobody comes here anymore, it's too crowded"; Predicting Image
  Popularity on Flickr}.
\newblock {\em Proceedings of International Conference on Multimedia Retrieval
  - ICMR '14}.

\bibitem[\protect\citeauthoryear{Meng \bgroup et al\mbox.\egroup
  }{2016}]{Meng:2016:MML:2946645.2946679}
Meng, X.; Bradley, J.; Yavuz, B.; Sparks, E.; Venkataraman, S.; Liu, D.;
  Freeman, J.; Tsai, D.; Amde, M.; Owen, S.; Xin, D.; Xin, R.; Franklin, M.~J.;
  Zadeh, R.; Zaharia, M.; and Talwalkar, A.
\newblock 2016.
\newblock Mllib: Machine learning in apache spark.
\newblock {\em J. Mach. Learn. Res.} 17(1):1235--1241.

\bibitem[\protect\citeauthoryear{Microsoft}{2017}]{TextAnalytics}
Microsoft.
\newblock 2017.
\newblock {Cognitive Services APIs} reference.
\newblock
  \url{https://westus.dev.cognitive.microsoft.com/docs/services/TextAnalytics.V2.0/operations/56f30ceeeda5650db055a3c9}.
\newblock Accessed: 2018-09-05.

\bibitem[\protect\citeauthoryear{Nesi \bgroup et al\mbox.\egroup
  }{2018}]{Nesi2018}
Nesi, P.; Pantaleo, G.; Paoli, I.; and Zaza, I.
\newblock 2018.
\newblock {Assessing the reTweet proneness of tweets: predictive models for
  retweeting}.
\newblock {\em Multimedia Tools and Applications}.

\bibitem[\protect\citeauthoryear{Palovics, Daroczy, and
  Benczur}{2013}]{Palovics2013a}
Palovics, R.; Daroczy, B.; and Benczur, A.~A.
\newblock 2013.
\newblock {Temporal prediction of retweet count}.
\newblock In {\em 4th IEEE International Conference on Cognitive
  Infocommunications, CogInfoCom 2013 - Proceedings}.

\bibitem[\protect\citeauthoryear{Peng \bgroup et al\mbox.\egroup
  }{2011}]{Peng2011}
Peng, H.~K.; Zhu, J.; Piao, D.; Yan, R.; and Zhang, Y.
\newblock 2011.
\newblock {Retweet modeling using conditional random fielDs}.
\newblock In {\em Proceedings - IEEE International Conference on Data Mining,
  ICDM}.

\bibitem[\protect\citeauthoryear{Peng \bgroup et al\mbox.\egroup
  }{2018}]{Peng2018}
Peng, S.; Zhou, Y.; Cao, L.; Yu, S.; Niu, J.; and Jia, W.
\newblock 2018.
\newblock {Influence analysis in social networks: A survey}.

\bibitem[\protect\citeauthoryear{Pezzoni \bgroup et al\mbox.\egroup
  }{2013}]{Pezzoni2013}
Pezzoni, F.; An, J.; Passarella, A.; Crowcroft, J.; and Conti, M.
\newblock 2013.
\newblock {Why do I retweet it? An information propagation model for
  microblogs}.
\newblock In {\em Lecture Notes in Computer Science (including subseries
  Lecture Notes in Artificial Intelligence and Lecture Notes in
  Bioinformatics)}.

\bibitem[\protect\citeauthoryear{Samek, Wiegand, and
  M{\"{u}}ller}{2017}]{Samek2017}
Samek, W.; Wiegand, T.; and M{\"{u}}ller, K.-R.
\newblock 2017.
\newblock {Explainable Artificial Intelligence: Understanding, Visualizing and
  Interpreting Deep Learning Models}.

\bibitem[\protect\citeauthoryear{Sapountzi and Psannis}{2018}]{Sapountzi2018}
Sapountzi, A., and Psannis, K.~E.
\newblock 2018.
\newblock {Social networking data analysis tools {\&} challenges}.
\newblock {\em Future Generation Computer Systems}.

\bibitem[\protect\citeauthoryear{Sheela}{2016}]{Sheela2016}
Sheela, L.~J.
\newblock 2016.
\newblock {A Review of Sentiment Analysis in Twitter Data Using Hadoop}.
\newblock {\em International Journal of Database Theory and Application}.

\bibitem[\protect\citeauthoryear{Tan, Lee, and Pang}{2014}]{tan2014effect}
Tan, C.; Lee, L.; and Pang, B.
\newblock 2014.
\newblock The effect of wording on message propagation: Topic- and
  author-controlled natural experiments on twitter.
\newblock In {\em Proceedings of the 52nd Annual Meeting of the Association for
  Computational Linguistics (Volume 1: Long Papers)},  175--185.
\newblock Baltimore, Maryland: Association for Computational Linguistics.

\bibitem[\protect\citeauthoryear{Wang, Bansal, and Frahm}{2018}]{Wang2018a}
Wang, K.; Bansal, M.; and Frahm, J.~M.
\newblock 2018.
\newblock {Retweet wars: Tweet popularity prediction via dynamic multimodal
  regression}.
\newblock In {\em Proceedings - 2018 IEEE Winter Conference on Applications of
  Computer Vision, WACV 2018}.

\bibitem[\protect\citeauthoryear{Wu and Shen}{2015}]{Wu2015}
Wu, B., and Shen, H.
\newblock 2015.
\newblock {Analyzing and predicting news popularity on Twitter}.
\newblock {\em International Journal of Information Management}.

\bibitem[\protect\citeauthoryear{Zaman \bgroup et al\mbox.\egroup
  }{2010}]{Zaman2010}
Zaman, T.~R.; Herbrich, R.; van Gael, J.; and Stern, D.
\newblock 2010.
\newblock {Predicting Information Spreading in Twitter}.
\newblock In {\em Workshop on Computational Social Science and the Wisdom of
  Crowds, NIPS 2010}.

\bibitem[\protect\citeauthoryear{Zhang, Si, and Hsieh}{2017}]{Zhang2017}
Zhang, H.; Si, S.; and Hsieh, C.-J.
\newblock 2017.
\newblock {GPU-acceleration for Large-scale Tree Boosting}.

\bibitem[\protect\citeauthoryear{Zhao \bgroup et al\mbox.\egroup
  }{2015}]{DBLP:journals/corr/ZhaoEHRL15}
Zhao, Q.; Erdogdu, M.~A.; He, H.~Y.; Rajaraman, A.; and Leskovec, J.
\newblock 2015.
\newblock {SEISMIC:} {A} self-exciting point process model for predicting tweet
  popularity.
\newblock {\em CoRR} abs/1506.02594.

\end{thebibliography}

\end{document}